\documentclass[conference]{IEEEtran}
\IEEEoverridecommandlockouts

\usepackage{float}
\usepackage{url}

\usepackage{multirow}

\usepackage{cite}
\usepackage{amsmath,amssymb,amsfonts}
\usepackage{algorithmic}
\usepackage{graphicx}
\usepackage{textcomp}
\usepackage{xcolor}
\def\BibTeX{{\rm B\kern-.05em{\sc i\kern-.025em b}\kern-.08em
    T\kern-.1667em\lower.7ex\hbox{E}\kern-.125emX}}
\begin{document}

\title{Beyond Quantile Methods: Improved Top-K Threshold Estimation for Traditional and Learned Sparse Indexes\\
}

\author{\IEEEauthorblockN{Jinrui Gou}
\IEEEauthorblockA{
\textit{New York University}\\
New York, US \\
jg6226@nyu.edu}
\and
\IEEEauthorblockN{Yifan Liu}
\IEEEauthorblockA{
\textit{New York University}\\
New York, US \\
yl8690@nyu.edu}
\and
\IEEEauthorblockN{Minghao Shao}
\IEEEauthorblockA{
\textit{New York University}\\
New York, US \\
ms12416@nyu.edu}
\and
\IEEEauthorblockN{Torsten Suel}
\IEEEauthorblockA{
\textit{New York University}\\
New York, US \\
torsten.suel@nyu.edu}
}

\maketitle

\begin{abstract}
Top-\textit{k} threshold estimation is the problem of estimating the score of the \textit{k}-th highest ranking result of a search query. A good estimate can be used to speed up many common top-\textit{k} query processing algorithms, and thus a number of researchers have recently studied the problem. Among the various approaches that have been proposed, quantile methods appear to give the best estimates overall at modest computational costs, followed by sampling-based methods in certain cases. In this paper, we make two main contributions. First, we study how to get even better estimates than the state of the art. Starting from quantile-based methods, we propose a series of enhancements that give improved estimates in terms of the commonly used mean under-prediction fraction (MUF). Second, we study the threshold estimation problem on recently proposed learned sparse index structures, showing that our methods also work well for these cases. Our best methods substantially narrow the gap between the state of the art and the ideal MUF of $1.0$, at some additional cost in time and space.
\end{abstract}


\section{Introduction}

Large search engines spend significant resources on processing user queries, motivating research on improved query processing methods that reduce this cost. Many search systems perform query processing using a cascade ranking approach \cite{lidanwang:cascadingsigir11}, where initially a fairly simple ranking function is used to select a large number of candidate results that are then reranked using more complex and expensive rankers. This candidate selection phase is commonly modeled as a disjunctive top-$k$ query, defined as follows: Given a document collection and associated index structures, a (fairly simple) scoring function $r$, and a query $q$, retrieve the $k$ documents with the highest scores. Widely studied optimized algorithms for disjunctive top-$k$ queries including MaxScore \cite{MaxScore}, WAND \cite{WAND}, and Block-Max methods such as \cite{BMW,dns:wsdm13,Carvalho-JIDM15}. 

In this paper, we consider a closely related problem called top-$k$ threshold estimation, where the goal is to estimate the score of the $k$-th highest scoring document for a given disjunctive top-$k$ query. Ideally, the estimate should be as tight as possible, use a limited amount of space, and be much faster than running the disjunctive top-$k$ query. In many scenarios, we also want to avoid overestimates. 

Threshold estimations can improve query processing efficiency in several scenarios. First, it can speed up many disjunctive top-$k$ query processing algorithms that use a top-$k$ threshold to avoid evaluating documents that cannot make it into the top $k$ results. Examples are MaxScore \cite{MaxScore}, WAND \cite{WAND}, and Block-Max methods \cite{BMW,dns:wsdm13,VBMW}. Without threshold estimation, such methods maintain a threshold estimate based on the results seen so far, initially set to $0$; this results in slow performance at the start of a query until enough documents have been encountered. It is well known that significant speed-ups can be obtained if we provide a good initial threshold estimate to these methods \cite{BMW,Carvalho-JIDM15,Petri-SIGIR19,Kane-SIGIR18,dns:sigir2013,threshold}
Second, threshold estimation can be used for resource selection in distributed search architectures such as selective search \cite{kulkarni2015selective} and index tiering \cite{risvik:latin,Leung,hauff}, where it allows us to estimate how many top-$k$ results are located in each shard or tier \cite{Aly-SIGIR13}.

\subsection{Threshold Estimation Methods}

This has motivated several recent studies of threshold estimation techniques \cite{Aly-SIGIR13,altingovde-cikm19,Petri-SIGIR19,sushi,threshold}, which can be grouped as follows: (1) parametric techniques, such as Taily \cite{Aly-SIGIR13}, that assume that document scores follow certain distributions, (2) ML-based techniques that use Bayesian Linear Regression or multi-layer perceptrons \cite{Petri-SIGIR19}, (3) sampling-based methods that estimate a threshold based on a small sample of the collection \cite{sushi,threshold}, and (4) quantile methods that precompute and store top-$k$ threshold values for terms and commonly occurring subsets of terms, and use these to lower-bound the thresholds of queries containing these terms or subsets \cite{altingovde-cikm19,threshold}.

An experimental study in \cite{threshold} showed that quantile and sampling-based methods significantly outperform the others in accuracy. In addition, quantile methods are also very fast, require limited space, and are {\em safe}\/ in that they never overestimate the real threshold. The latter is desirable in top-$k$ query processing algorithms, since an overestimate would require re-execution of the query with a lower estimate in order to guarantee the correct top-$k$ results. Sampling-based methods can get similar accuracy as quantile methods, though with slightly higher computing and space overheads, and allow us to trade-off estimation accuracy against speed, space, and the likelihood of an overestimate. 

\subsection{Our Contributions}

In this paper, we propose and evaluate new threshold estimation methods with improved accuracy over existing methods. In particular, our contributions are as follows:
\begin{itemize}
\item[(1)]
We describe new threshold estimation techniques that build on top of quantile methods, and augment them with ideas from top-$k$ query processing methods \cite{fagin,candi}.
\item[(2)] 
We provide an extensive experimental evaluation shows that our methods can substantially narrow the remaining gap between estimation methods and the precise top-$k$ threshold. In particular, our methods significantly improve estimation for longer queries, a case for which existing quantile methods do not perform well. 
\item[(3)]
We provide the first study of threshold estimation techniques for learned sparse index structures. The results show that for indexes obtained with DocT5Query document expansion \cite{docT5query} and with DeepImpact \cite{deepimpact2021}, our methods still perform well, though there are some interesting differences with the case of traditional indexes.
\end{itemize}
\section{Background and Related Work}

We now define the problem, describe the best previous techniques, and briefly discuss other closely related work. 

{\bf Problem Definition:}
The threshold estimation problem is defined as follows: Given a set of documents $D$, a query $q$, an integer $k$, and a scoring function $sc$ assigning each document $d \in D$ a score $sc(q,d)$ with respect to $q$, estimate the score of the $k$-th highest scoring document in $D$. As in previous work, we assume that the scoring function has the form $sc(q,d) = \sum_{t \in q}  sc(t, d)$ where $sc(t, d)$ is the score of $d$ with respect to term $t$. Thus the term scores $sc(t,d)$, also called term impact scores, can be precomputed during indexing and stored, say, as postings of a quantized inverted index. 

The accuracy of threshold estimation is evaluated with two measures, the \textit{Mean Under prediction Fraction} (MUF) measure proposed in \cite{Petri-SIGIR19}, defined as the ratio of the estimated and the real top-$k$ threshold, averaged over all queries that do not result in an overestimate, and the rate of overestimates, which is zero for quantile methods and most of the methods presented here.

{\bf Quantile Methods:} To describe existing quantile-based methods, We start with single-term quantile methods where, given a $k$, say $k$=$10$ or $1000$, we store for each term $t$ in the collection the $k$-th highest value of term impact scores $sc(t,d)$, over all $d \in D$, called $th(t, k)$. For a query $q$, we provide a lower-bound estimate for the top-$k$ threshold of $q$ by computing $\max_{t \in q} th(t, k)$. 

This idea can be extended to multi-term quantile methods where we choose subsets of two, three, or even four terms that frequently occur together in queries. For each such subset $s$, we store its top-$k$ threshold, $th(s,k)$, defined as 
the $k$-th highest value of $\sum_{t \in s} sc(t,d)$ over all $d \in D$. When a query $q$ arrives, we estimate the top-$k$ threshold of $q$ by computing $\max_{s \subseteq q} th(s, k)$. This is clearly a lower bound of the actual threshold. Moreover, we would expect a tighter estimate from larger subsets that are contained in the query. The computational cost of quantile-based threshold estimation is dominated by the cost of looking up all subsets of the query for which quantiles have been precomputed and stored.

The main challenge for multi-term quantile methods is to select subsets $s$ for which we precompute and store $th(s, k)$, as it is infeasible to do this for all subsets of $2$, $3$ or more terms. The study in \cite{threshold} compared two methods, a lexical one that chooses subsets that frequently co-occur in documents, and a log-based one that chooses subsets that frequently co-occur in queries from a large log. The log-based approach performed better, and we will use it here. 

{\bf Random Sampling:} The idea in the sampling-based approach \cite{sushi,kulkarni2012shard,threshold} is to calculate or estimate the top-$k'$ threshold on a smaller, sampled set of documents, and use it to estimate the top-$k$ threshold for the entire set, for an appropriately chosen $k' < k$. The overestimation rate, denoted by $O$, can be derived as: $
O = \sum_{i=k'}^{k-1} \binom{k-1}{i} \cdot s^i \cdot (1 - s)^{k-i-1}$, where $s$ represents the sample rate such that, e.g., for $s=0.1$ each document would be included in the sample independently with probability $0.1$. Choosing a value of $k'$ close to $s\cdot k$ results in a better MUF but a higher rate of overestimates, while a slightly larger $k'$ can significantly reduce the overestimation rate. In \cite{threshold}, sampling rates between $0.002$ and $0.01$ were used, and thresholds on the sample were computed by running a safe disjunctive top-$k'$ algorithm. In our approach, we instead use our new estimation methods directly on the sample, allowing us to use higher sample rates ($0.02$ to $0.05$) without an unacceptable loss in efficiency. 

{\bf Top-$k$ Query Processing Algorithms:} Our work is related to top-$k$ query processing algorithms in two ways. First, the main application of threshold estimation is to improve the efficiency of such algorithms. For example, MaxScore-based methods \cite{MaxScore,michal:wsdm22} use thresholds to select non-essential terms that do not need to be completely traverse, while WAND-based methods \cite{WAND,magicofwand,BMW} use them to select pivot docIDs, and block-filtering methods \cite{dns:sigir2013,mallia:wsdm21} use them to skip parts of the docID space.
Second, our approach in this work in turn applies ideas from existing query processing algorithms to the threshold problem. In particular, our approach is influenced by the unsafe top-$k$ processing approach in \cite{candi}, which itself builds on the TA method for safe top-$k$ queries in \cite{fagin}.

{\bf Sparse Learned Indexes:} Recent advances in Transformers and Large Language Models have motivated a number of proposals for creating Learned Sparse Indexes, i.e., inverted index structures obtained by using these powerful models to derive suitable index postings and term impact scores. The goal is to improve result quality while retaining the performance advantages of inverted indexes. Examples of such approaches are COIL \cite{coil} and uniCOIL \cite{unicoil}, SPLADE \cite{splade:sigir2021}, and techniques based on document expansion such as DocT5Query \cite{docT5query} and DeepImpact \cite{deepimpact2021}. We focus here on the latter two, as the former have query-dependent term weights that require additional ideas to make our approach work.

It has been observed that these approaches result in inverted index structures with score distributions that are quite different from those in standard inverted indexes \cite{deepimpact2021,wackyweights}, leading to slower running times for well-known top-$k$ query processing algorithms such as MaxScore or BMW.  This in turn has motivated recent work on query processing methods that work well with these new indexes, such as \cite{guidedtraversal,ioqp,termimpactdecomposition}. Threshold estimation techniques can be a useful tool for faster query processing, but have to our knowledge not yet been tested on these indexes. We evaluate our methods on indexes based on DocT5Query document expansion and DeepImpact.

\section{Our Approach}

We are now ready to describe our new methods. We present these as a series of extensions starting from the quantile methods described in the previous section, which each aiming to provide a boost in estimation accuracy at a slight increase in processing cost or space. However, the complete final method can also be seen as a very aggressive early termination technique for disjunctive top-$k$ queries based on the general approach in \cite{candi}, but used for threshold estimation.

{\bf Using Several Subsets via Duplicate Removal:} One limitation of quantile methods is that the estimate is determined by a single threshold $th(s,k)$ for the particular subset $s$ that provides the best estimate for this query. Now consider a query $q$ and two subsets $s_1$ and $s_2$ of $q$ with $th(s_1, 10) = 10.0$ and $th(s_2, 10) = 9.0$. Thus, we return $10.0$ as the best estimate. But suppose we know that $th(s_1, 6) = 12.0$ and $th(s_2, 4) = 12.5$ -- could we can return a better top-$10$ estimate of $12.0$ since there are at least $4+6 = 10$ documents scoring at least $12.0$? This might get better estimates by storing thresholds for additional values of $k$.

However, the $6$ highest scoring documents for $s_1$ and the $4$ highest scoring documents for $s_2$ might not be disjoint. To get a valid lower-bound estimate, we would have to detect any duplicates and remove them from consideration. To do so, we would have to store not just additional threshold values for other values of $k$, but for each selected subset $s$ we also need a prefix of the docIDs of the highest scoring documents with respect to $s$. This clearly increases storage costs, especially for higher values of $k$, but might be worth it if it gives a significant boost in accuracy. We refer to this method as {\em Remove Duplicates}, defined as follows:

{\em Remove Duplicates: At indexing time, for each selected subset $s$, we store a sorted prefix of the $k$ highest-scoring results of a disjunctive query for $s$, where each result has a score and a docID. To get a threshold estimate for a query $q$, we identify all subsets that are available, and then select docIDs from the stored prefixes of these subsets in decreasing order of score, until $k$ distinct docIDs have been retrieved, and return the score of the $k$-th docID as the estimate.}

We observe that if we can afford to store such prefixes for all subsets used by the quantile method, then we are guaranteed to get a safe estimate (i.e., with no overestimates) that is at least as good as that provided by the quantile method.

{\bf Combining Scores from Different Structures:} The next natural step is to combine any term scores retrieved from different prefixes that have the same docID. However, we have to be careful about how to do this.

For example, assume we need a top-$10$ threshold estimate for $q = \{x, y, z\}$, and have precomputed and stored threshold data for the subsets $\{x\}$, $\{y\}$, $\{x,y\}$ and $\{y, z\}$ of $q$. Suppose that for each such subset, we have stored a prefix with the posting scores and docIDs for the say $50$ highest scoring documents. Then we could try to get a better estimate by fetching the docID and term score information for all $50$ highest scoring documents for the $4$ subsets, and combining the term scores for any docIDs discovered in more than one subset. However, this assumes that we store for each docID in the prefix of a subset not just the total score, but each constituent term score, to avoid adding up the same term score more than once. Also, we may want to store prefixes of more than $k$ postings for our subsets, which requires that subsets and prefix depths have to be selected for maximum utility. We refer to this method as {\em Combine Scores}, defined as follows:

{\em Combine Scores: Based on query log analysis, we select suitable subsets and associated prefix lengths. At indexing time, we store for each subset $s$ a sorted prefix of the highest scoring results of a conjunctive query on the subset terms, up to the chosen prefix length, where each result consists of a docID and the scores of all terms in $s$. To get a threshold estimate for a query $q$, we identify all subsets of $q$ that are available, and then select postings from the stored prefixes of these subsets in decreasing order of total score, until a certain number of postings have been processed, as determined by a access budget $ab$. While processing postings, we combine scores from different structures with the same docID, using an accumulator data structure such as a hash table. We return the $k$-highest result in the hash table as our estimate.}

This method is again safe, and returns an estimate at least as good as that of {\em Remove Duplicates} if we access enough postings to get $k$ distinct docIDs. The reason for now storing top results from a conjunctive rather than a disjunctive query in the prefix is to avoid storing redundant entries in different prefixes. Suppose the prefix stores the top-10 results of a disjunctive query on subset $\{t_1, t_2, t_3\}$, and the highest-scoring entry has term scores $sc(t_1, d) = 128$, $sc(t_2, d) = 0$ and $sc(t_3, d) = 64$. Thus, $t_2$ does not occur in document $d$, and in subset $\{t_1, t_3\}$, docID $d$ would also be the highest-scoring entry. In most cases, when a prefix for a subset $s$ is available, we also have prefixes available for all subsets of $s$, and thus the above entry in the prefix for $\{t_1, t_2, t_3\}$ would be redundant as we would find the same information in the prefix $\{t_1, t_3\}$ when postings are accessed by decreasing overall score. Thus, using a conjunctive query for building prefixes makes better use of the space budget.

{\bf Adding Lookups:} The next natural step is to also add index lookups to obtain missing term scores for the results accumulated in the hash table of the {\em Combine Scores}\/ method. In particular, we have a lookup budget $lb$, and perform lookups into an inverted index for the top $lb$ results returned by {\em Combine Scores}\/ for all unknown term scores. We refer to this method as {\em Lookups}. To get benefits, we require $lb>k$.

{\bf Discussion:} Note that our complete method with lookups is reminiscent of certain optimized top-$k$ query processing techniques in the literature that perform early termination by accessing postings in order of decreasing term scores, followed by lookups into an index to resolve any missing term scores, such as \cite{fagin,candi}. The main differences here are that we also explore intermediate approaches such as {\em Remove Duplicates}\/ and {\em Combine Scores}, and that we have more stringent time constraints, as determined by $ab$ and $lb$, as we need running times much lower than those for top-$k$ query processing.

We also note that our implementations, unless stated otherwise, are backed up by a state-of-the-art quantile method. That is, the estimates we return are the maximum of the estimate returned by our method and the quantile method. Because our method must store larger prefixes, it cannot cover as many subsets as a quantile method that needs to only store a single score per prefix. Thus, for some queries, the quantile method can outperform our method without backup.
Consequently, our methods are by definition as least as precise as the quantile method. To prove their usefulness, we need to show that they significantly outperform the quantile method under acceptable increases in CPU and space costs.

{\bf Adding Random Sampling:} Finally, we explore how to incorporate the sampling method from \cite{threshold} into our approach. As we will see, large values of $k$ require larger access and lookup budgets $ab$ and $lb$, and with it also the storage of large enough prefixes that allow us to actually use these budgets. Sampling basically reduces a top-$k$ threshold estimate to a top-$k'$ threshold estimate for a much smaller value of $k'$ on a sample of the collection. This reduces CPU and space costs, but also introduces an additional estimation error due to sampling and a small chance of an overestimate. Formally, the method is as follows:

{\em Adding Sampling: We select a sampling rate $s$, and build prefixes and an inverted index on the resulting sample of the collection. To get a top-$k$ threshold estimate for a query $q$, we select a value of $k'$ such that limits the probability of an overestimate to an acceptable level, say $0.01$\%. We then use our methods to get a top-$k'$ threshold estimate on the sample.}

\section{Experimental Results}

We now present a detailed experimental evaluation of our methods in terms of accuracy (as measured by MUF and, if applicable, overestimation rates), space requirements, and speed. Recall that the Mean Underestimation Rate (MUF) of a method is defined as the ratio of the top-$k$ threshold estimate provided by the method over the precise top-$k$ threshold, averaged over those queries where no overestimate occurs.

Our evaluation has two parts. We first evaluate the different methods and choices of prefix types and budget in a setting where we store very large prefix structures, with focus on BM25 as ranking function. This gives us a general idea of how various choices perform and what to focus on. Then we adopt the most promising choices, and study how to reduce space consumption to practical levels, how the method performs for different types of ranking functions, including sparse learned methods, and what the actual CPU costs are.


{\bf Setup:} All algorithms were implemented in C++17, and compiled with GCC 11.4.0 with -O3 optimization. Runs were performed on a single core of a 3.2GHz Intel Core i9-12900K CPU of a machine, running GNU/Linux 5.15. The index used for lookups was built with the PISA\cite{PISA2019} framework and compressed using Elias-Fano coding\cite{vigna2013elias} for fast lookup performance. ALLookups into each inverted list were performed in ascending order of docID, to maximize performance. All prefixes and indexes are in main memory. Prefixes were selected based on various criteria described in the experiments, and generated by issuing queries to the index.

As baseline for comparisons, we implemented the quantile method $Q^4_k$ from \cite{threshold}, which stores top-$k$ quantiles for subsets of up to $4$ terms. Subsets for $Q^4_k$ were selected based on a training query log, and all subsets occurring in the log were included. Unless stated otherwise, our methods are backed up by $Q^4_k$, i.e., the $Q^4_k$ estimate is returned when it is higher than that returned by our method. The $Q^4_k$ quantiles are stored in the same data structure that is used to fetch our prefixes, and thus adding $Q^4_k$ does not increase software complexity by much.

{\bf Datasets:} We ran experiments on several datasets, and present results for the ClueWeb09B document collection, consisting of about 50 million documents and 88 million distinct terms, and the MSMARCO Passage Ranking dataset \cite{nguyen:msmarco2016} of about 8.8 million passages. All terms were lowercased and stemmed by the Porter2 stemmer. Document IDs were assigned in lexicographic order of URLs. Unless stated otherwise, the scoring function is BM25. 

We also used the AOL query log as a training log to determine for which subsets to precompute and store quantile scores and prefixes of docIDs and term scores. We built prefixes for subsets up to size $4$ (singles, pairs, triplets, and quadruplets), but also explored limiting subsets to smaller sizes. For evaluation, we randomly chose 10k queries from the TREC 2005 Terabyte Track Efficiency Task data, excluding single term queries for which threshold estimation is trivial.

\subsection{Part I: Basic Choices and Tradeoffs}\label{subsec:basic_result}

We first evaluate some basic choices in our methods under the assumption of a very large space budget, for the BM25 scoring function. In particular, we build prefix structures up to certain depths for all single terms occurring in the index, and for all pairs, triplets, and quadruplets that appear in at least one query in the AOL training log. The depths are chosen as $10k$ for single terms and pairs, $4k$ for triplets and $3k$ for quadruplets. This means that for reasonable access budgets $ab$, we are unlikely to reach the cutoff point of such a structure. We also stored a top-$k$ quantile score for each subset that was selected, for use by the $Q^4_k$ method.

\begin{figure}[H]
\vspace{-3mm}
\centering
\includegraphics[width=0.77\columnwidth]{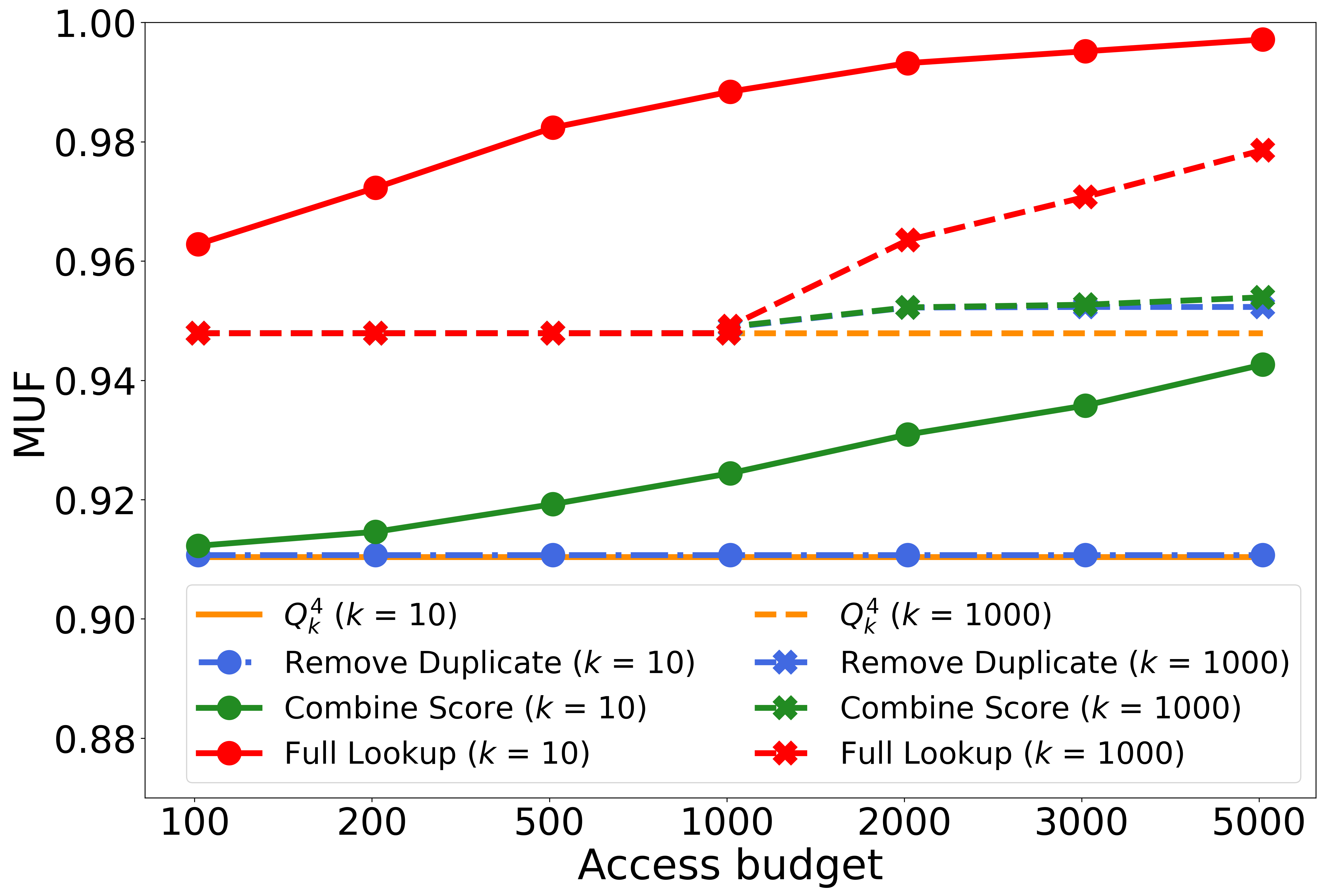}
\caption{Comparison of methods for \textit{k} = 10 and \textit{k} = 1000, for ClueWeb09B.}
\label{fig:compare_methods}
\vspace{-3mm}
\end{figure}

Figure~\ref{fig:compare_methods} compares $Q^4_k$ with our three newly proposed methods on ClueWeb09B data, for $k = 10$ and $k = 1000$ and with access budget $ab$ ranging from $100$ to $5000$. We performed full lookups for the third method, that is, $lb = ab$. 

We first discuss results for $k=10$, shown in solid lines. We see at the bottom of the chart that {\em Remove Duplicates}\/ provides basically no benefit over $Q^4_k$ for this case as the curves are on top of each other. This is not surprising as in many cases the top 10 postings are chosen from the same prefix, resulting in the same estimate as $Q^4_k$, while {\em Remove Duplicates}\/ can only benefit if several prefixes are involved. Also, as soon as $k$ distinct docIDs are encountered, additional access budget is not useful anymore. {\em Combine Scores}\/ does give noticeable improvements for $k-10$. However, these improvements are dwarfed by the huge benefit of performing lookups, where we see MUF values approaching $1.0$ for $k=10$. 

Looking at $k=1000$ in the dashed lines, we see minor benefits for {\em Remove Duplicates}\/ for $ab >1000$, the minimal possibly useful budget for $k=1000$; this is because the first $1000$ selected postings are less likely to come from just one prefix. However, {\em Combine Scores}\/ gives little benefit beyond this, while full lookups again give a big boost, though not as much as for $k=10$.
Results for MSMARCO (not shown) showed similar trends. Thus, while we initially hoped that eliminating duplicates and combining scores might give significant improvements over quantile methods, the results suggest that lookups are a crucial part of a successful approach.

\begin{figure}[H]
\vspace{-4mm}
\centering
\includegraphics[width=0.77\columnwidth]{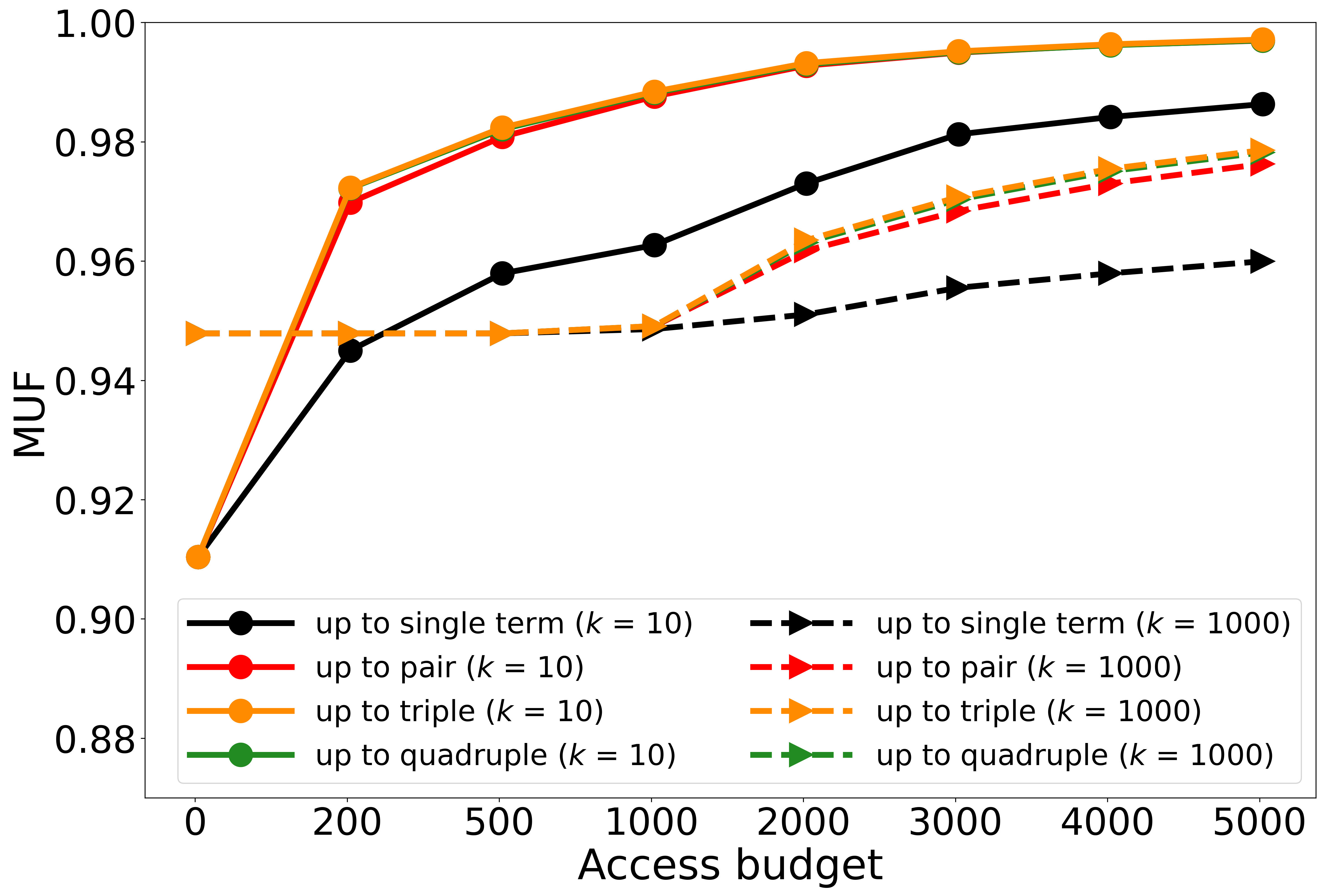}
\caption{Comparing singles, pairs, triples, and quadruples for ClueWeb09B.}
\label{fig:compare_add_grams_cw09b}
\vspace{-3mm}
\end{figure}

Figure~\ref{fig:compare_add_grams_cw09b} shows the impact of using pairs, triples, and quadruples in the proposed method, for the case of ClueWeb09B. We show $k=10$ in solid lines, $k=1000$ in dashed lines, and the performance of the $Q^4_k$ method as the point with access budget $0$ on the left. We perform full lookups on all encountered docIDs.
Adding quadruples provides no benefits over using triples, as their lines are completely on top of each other. Triples provide only minor benefits over pairs. For $k=10$, even singles provide significant benefits over $Q^4_k$, while the benefits are smaller for $k=1000$. In both cases, pairs provide a significant boost over singles.

\begin{figure}[H]
\vspace{-3mm}
\centering
\includegraphics[width=0.82\columnwidth]{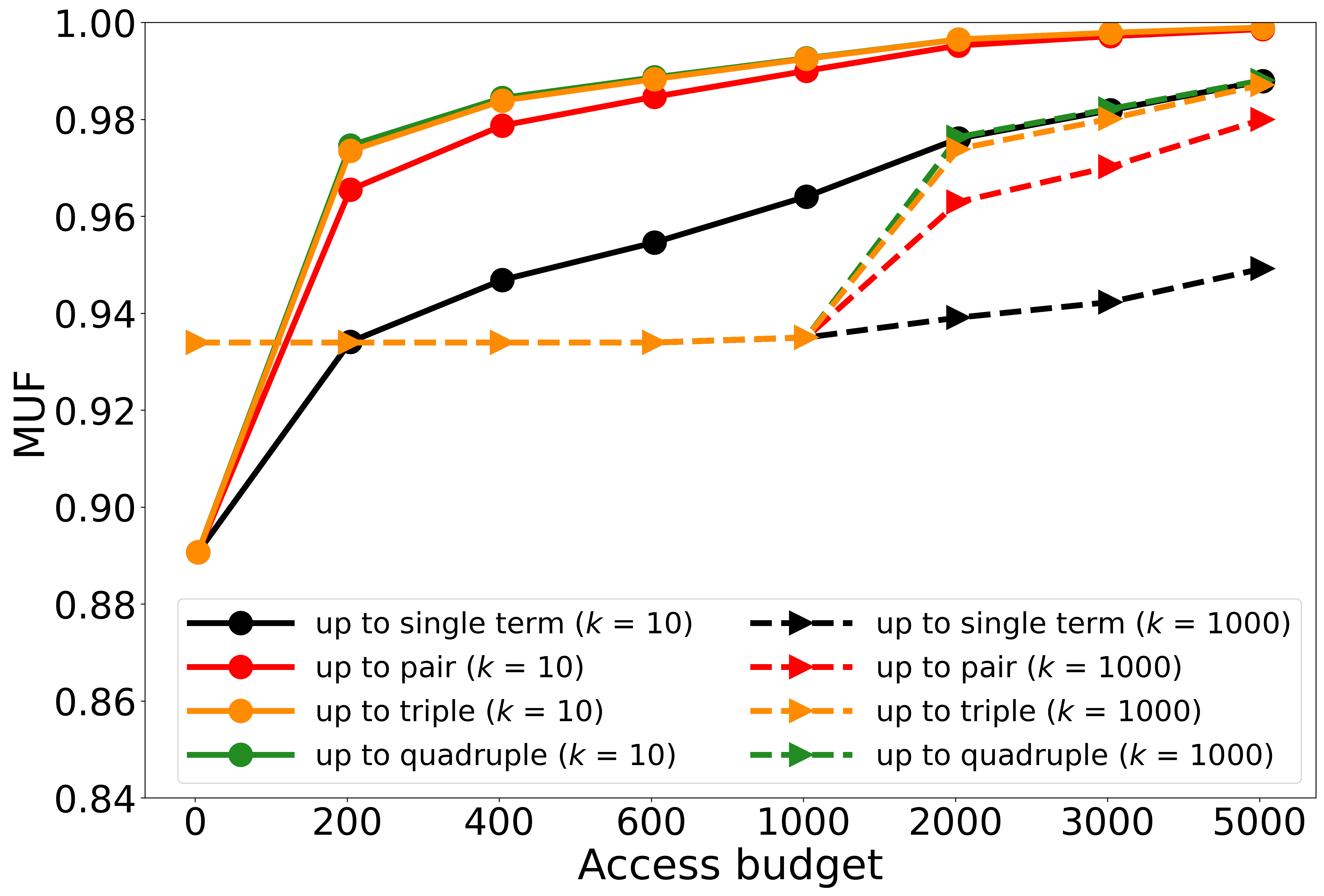}
\caption{Comparing singles, pairs, triples, and quadruples for MSMARCO.}
\label{fig:compare_add_grams_msmarco}
\vspace{-3mm}
\end{figure}

Results for MSMARCO in Figure \ref{fig:compare_add_grams_msmarco} show slightly more benefits for triples over pairs, but the general trend is similar. Overall, using pairs gives a significant boost, while triples and quadruples do not help much, especially for larger budgets.

\begin{figure}[H]
\vspace{-3mm}
\centering
\includegraphics[width=0.82\columnwidth]{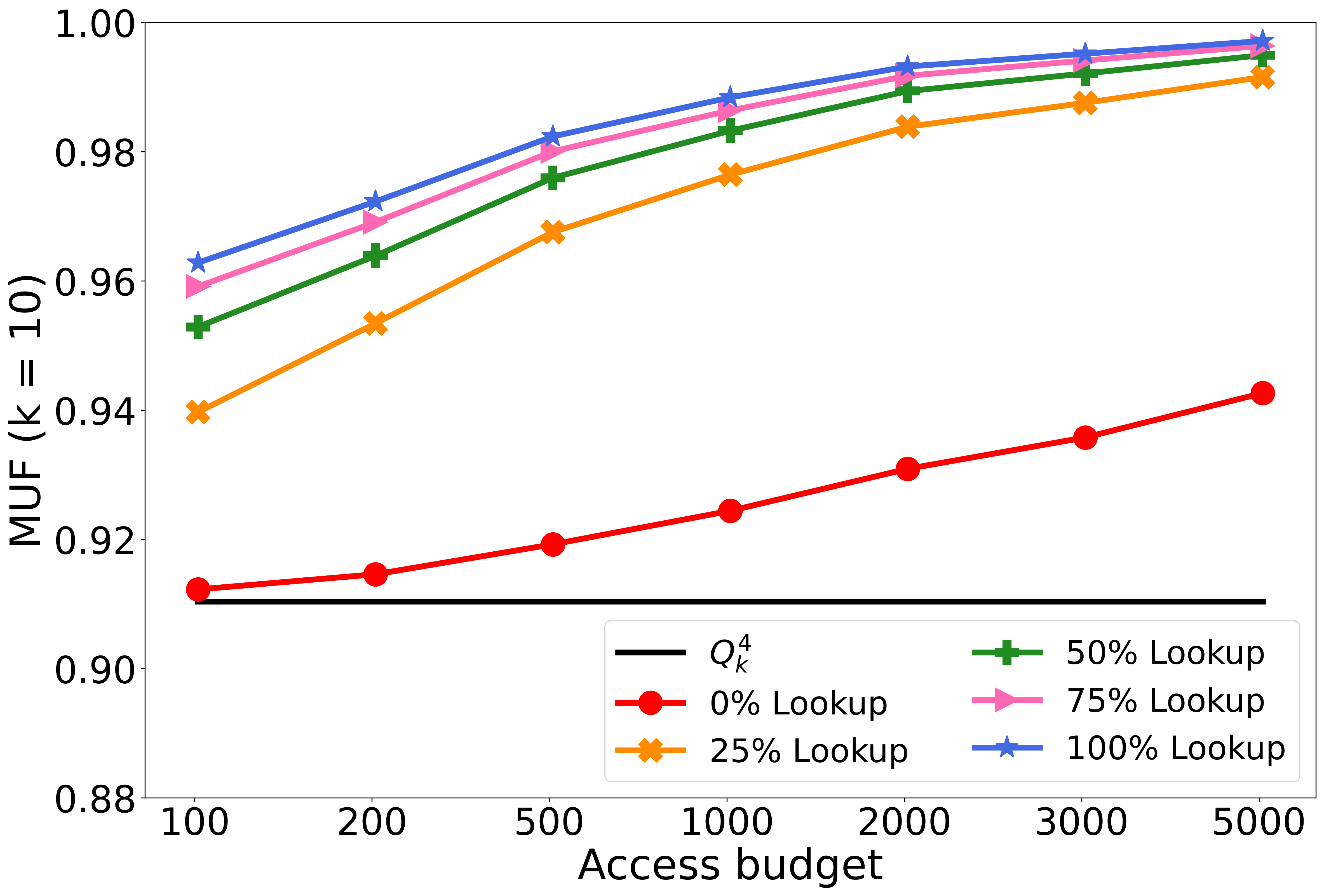}
\caption{MUF of different lookup ratios for \textit{k} = 10 on ClueWeb09B.}
\label{fig:compare_lookup_ratio}
\vspace{-3mm}
\end{figure}

Figure~\ref{fig:compare_lookup_ratio} looks at the lookup budget $lb$ as a fraction of the access budget $ab$, for $k=10$ on ClueWeb09B. Of course, $lb=ab$ performs best, but even $50\%$ gives decent results.

Table~\ref{tab:combined_lookup_ql} shows results for different query lengths on Clueweb09a, for full lookups, For short queries improvements over $Q^4_k$ are minor, but for longer queries $Q^4_k$ significantly deteriorates, while our methods still do well. 

\begin{table}[htbp]
  \vspace{-4mm}
  \caption{MUF for different access budgets (\textit{ab}) and query lengths.}
  \centering
\vspace{-3mm}
  \scriptsize
  \begin{tabular}{ccccccc}
      & 2 & 3 & 4 & 5 & 6+ & avg \\
    \hline
    \# Queries& 3725 & 2726 & 1690 & 893 & 966 &  \\
    \hline
    \multicolumn{7}{l}{\textbf{\textit{k} = 10}} \\
    $Q_k^4\textit{-log}$ & 0.960 & 0.923 & 0.884 & 0.851 & 0.786 & 0.910 \\
    \textit{ab} = 200 & 0.980 & 0.975 & 0.968 & 0.962 & 0.953 & 0.972 \\
    \textit{ab} = 500 & 0.986 & 0.985 & 0.980 & 0.976 & 0.971 & 0.982 \\
    \hline
    \multicolumn{7}{l}{\textbf{\textit{k} = 1000}} \\
    $Q_k^4\textit{-log}$ & 0.984 & 0.959 & 0.933 & 0.904 & 0.844 & 0.948 \\
    \textit{ab} = 2000 & 0.988 & 0.971 & 0.949 & 0.927 & 0.888 & 0.962 \\
    \textit{ab} = 5000 & 0.991 & 0.981 & 0.974 & 0.964 & 0.946 & 0.978 \\
    \hline
  \end{tabular}
  \vspace{-2mm}
  \label{tab:combined_lookup_ql}
\end{table}


{\bf Summary of Part I:} The main observations from our first set of experiments are: (1) lookups are very important for estimation quality; (2) using prefixes for pairs gives a significant boost over single terms while triples and quadruples give limited benefit; (3) lookup budgets can be limited to a subset of encountered docIDs without much degradation; and (4) our methods do particularly well on long queries, while for short queries quantile methods are sufficient.

\subsection{Part II: Space, Ranking Functions, and Speed}\label{subsec:more_result}

Drawing from the lessons of the first part, we now explore additional aspects of our methods. We first address the space issue, and propose more realistic configurations that still achieve decent benefits. Then we look at different ranking functions, including those used in learned sparse indexing methods. Finally, we look at the run times of our methods and their impact when used to accelerate the MaxScore algorithm.

\begin{figure}[H]
\vspace{-3mm}
\centering
\includegraphics[width=0.85\columnwidth]{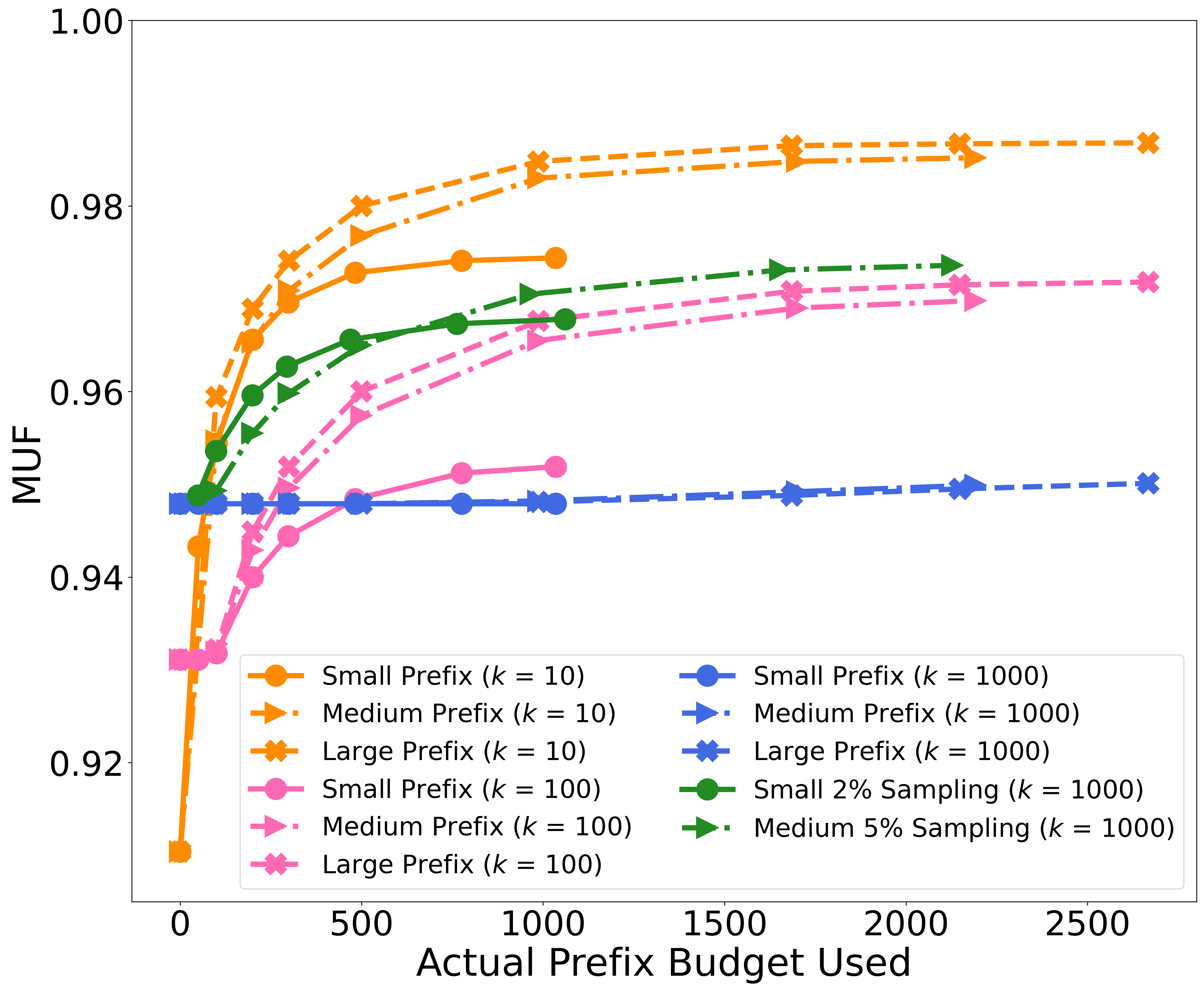}
\caption{MUF of different prefix configurations for \textit{k} = 10, 100, 1000.}
\label{fig:vary_prefix_size}
\vspace{-3mm}
\end{figure}

{\bf Space Management:} 
Our experiments so far have used impractically large prefix structures, as we created prefixes for all subsets up to size $4$ that occurred in a large query log. This resulted in about $1704$ GB space for storing prefixes for the ClueWeb09B collection. To make the methods practical, we need to reduce this size to an amount comparable to, or preferably lower than, the index size, which is $25.6$ GB. 

We now show results for three configurations on the ClueWeb09B dataset that are all smaller than the index. This achieve this, we get rid of subsets of size $4$, severely cut back on subsets of size $3$, and choose the prefix sizes of the remaining subsets adaptively depending on how frequent the subsets are in the training query log. In particular, we constructed a small prefix ($2.7$ GB) that only stores a small number of postings for single terms and pairs occurring in the AOL training log, a medium prefix ($6.8$ GB) that keeps a few more postings than the small prefix for each prefix, and a large prefix ($14.97$ GB) that also adds a small number of prefixes for triplets that occur frequently in the AOL training log. Precise definitions of these configurations will be made available as part of the dataset release. 

During testing, we observed that in many cases it  was impossible to fully use the available access budget, particularly for the small and medium configurations, as there were not enough relevant prefixes available for many queries. This problem had not occurred much before on the very large structures used in Part I, where available and actually used budgets were very close. We thus decided to plot the actual access budget used on the x-axis, as this is a more accurate measure of computational cost.

Results are in Figure~\ref{fig:vary_prefix_size}, for the three prefix configurations and for $k = 10$, $100$, and $1000$. We perform full lookups. As before, we plotted $Q^4_k$ as the leftmost point with access budget $0$. We selected access budgets up to $5000$, but plotted the actual budget that was used on the x-axis -- if a graph ends early on this axis, it means that any larger access budgets could not be fully used for the given prefixes.

For $k =10$ and $k=100$, our method still significantly outperforms $Q^4_k$ on all three prefixes. However, for $k=1000$ (blue lines), we do not see any benefits over $Q^4_k$. The reason is partially the limited access budget, but more importantly the fact that the limited access budget is not even fully used given the very short prefix structures. However, when we add sampling, for $s=0.02$ and $s=0.05$ with underestimation limited to $0.01\%$, we again see significant benefits for our methods. Thus, large values of $k$ require larger prefixes, unless sampling is used to reduce $k$ to a smaller $k'$.

{\bf Exploring Different Ranking Functions:} Next, we look at the performance of our methods on different ranking functions, including functions used in learned sparse indexing approaches. In these experiments, we focus on MSMARCO, as this was the only dataset for which all four indexes were available to us. We use the following ranking functions: (1) BM25 on the original index, as used before, (2) QLD, which is Query Likelihood with Dirichlet smoothing on the original index \cite{ponte1998language}, (3) BM25 on an index where documents have been expanded with DocT5Query \cite{docT5query}, and (4) DeepImpact \cite{deepimpact2021} based on DocT5Query document expansion. 
In the next two figures, we again use index structures that contain prefixes for all subsets of size up to $4$ that occur in the training queries.

\begin{figure}[H]
\vspace{-4mm}
\centering
\includegraphics[width=\columnwidth]{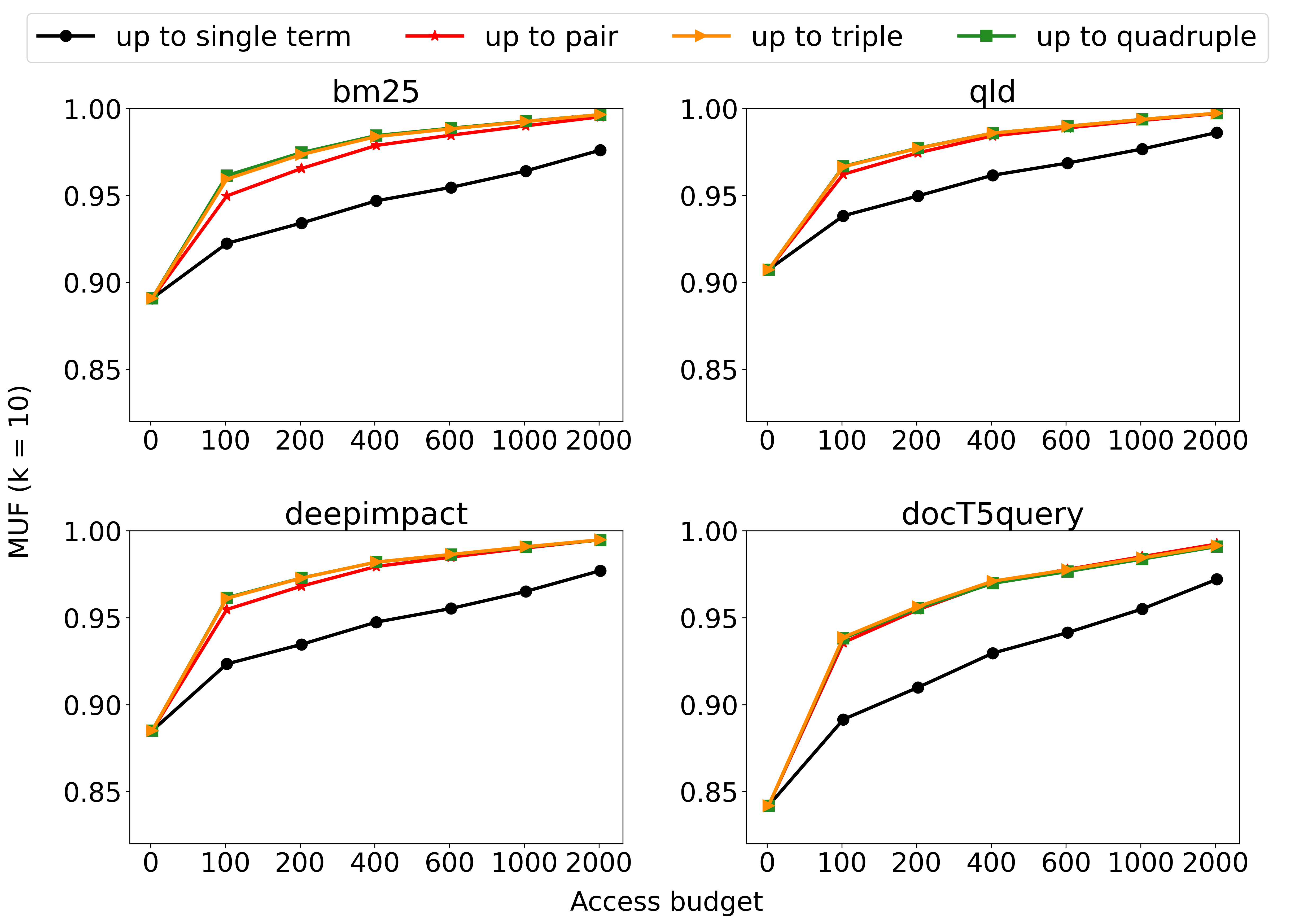}
\caption{Comparison of singles, pairs, triples, and quadruples for \textit{k} = 10 on MSMARCO. We compare four different ranking methods, BM25, QLD, a DocT5Query expanded index, and DeepImpact.}
\label{fig:compare_methods_msmarco_top10}
\vspace{-3mm}
\end{figure}
Figure~\ref{fig:compare_methods_msmarco_top10} shows results for $k=10$, for all ranking functions and different subset sizes. As for BM25 on ClueWeb09B, we observe good benefits over $Q^4_k$ for single terms, much greater benefits for pairs, but little benefits for triples and quadruples. 

\begin{figure}[H]
\vspace{-3mm}
\centering
\includegraphics[width=\columnwidth]{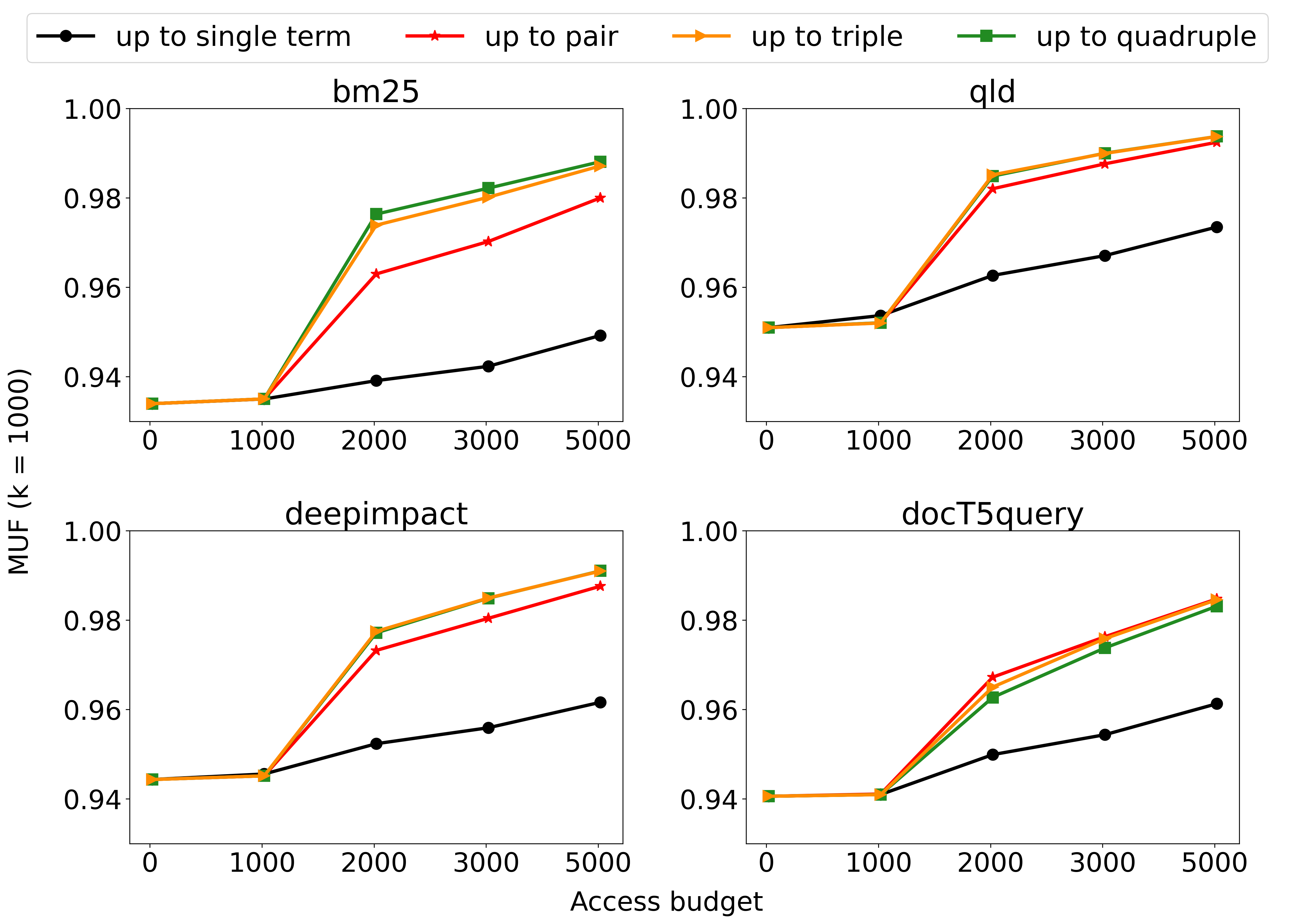}
\caption{Comparison of singles, pairs, triples, and quadruples for \textit{k} = 1000 on MSMARCO. We compare four different ranking methods, BM25, QLD, a DocT5Query expanded index, and DeepImpact.}
\label{fig:compare_methods_msmarco_top1000}
\vspace{-3mm}
\end{figure}

Figure~\ref{fig:compare_methods_msmarco_top1000} shows results for $k=1000$. We see slightly more benefit for triplets over pairs for BM25 and DeepImpact. Overall, we see that our methods achieve significant improvements in estimation quality over $Q^4_k$ (plotted as the leftmost point with access budget $0$) for all ranking functions. QLD in particular does very well on MUF, but the result for DeepImpact is also surprising, given that its score distributions deviate significantly from BM25.

\begin{figure}[H]
\centering
\vspace{-3mm}
\includegraphics[width=0.85\columnwidth]{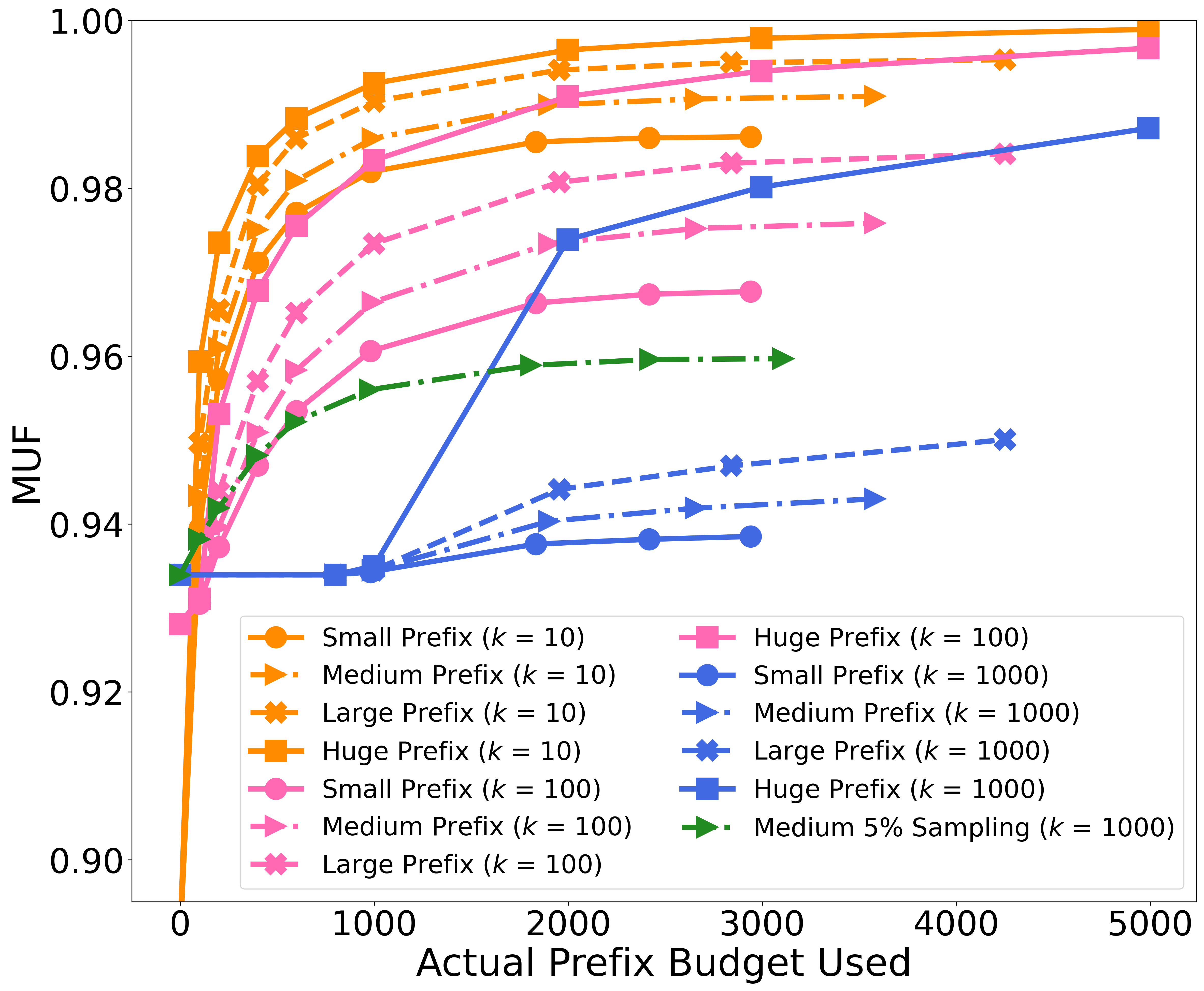}
\caption{MUF of different prefix configurations for \textit{k} = 10, 100, 1000 on MSMARCO, using BM25 as ranking function. For k = 10, the MUF returned by $Q_k^4$ is 0.891, which is not shown in the figure for better readability.}
\label{fig:vary_prefix_size_msmarco_bm25}
\vspace{-4mm}
\end{figure}

Next, we evaluate how the different ranking functions fare when we restrict the sizes of the prefix structures to more reasonable values. We start out with BM25 on MSMARCO, to set a baseline for the other ranking functions. Figure \ref{fig:vary_prefix_size_msmarco_bm25} shows results for $k=10$, $100$, and $1000$, for four different prefix configurations: the small, medium, and large configurations from before, and the huge configuration from Part I. We again see similar behavior as for ClueWeb09B, with good performance for $k=10$ and $k=100$, little benefit for $k=1000$ without sampling, and again good benefit once sampling is integrated into our approaches.

\begin{figure}[H]
\centering
\vspace{-3mm}
\includegraphics[width=0.82\columnwidth]{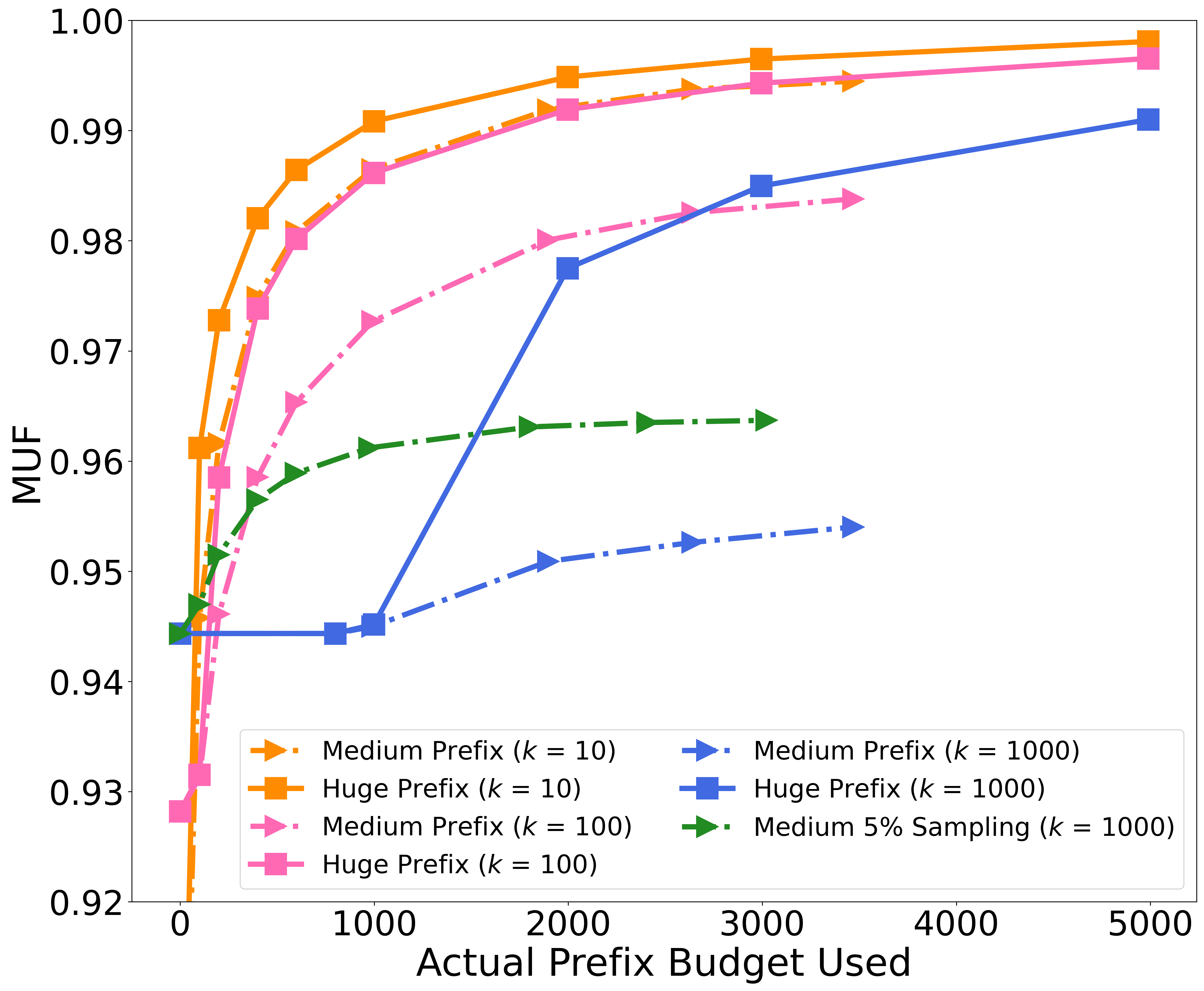}
\caption{MUF of different prefix configurations for \textit{k} = 10, 100, 1000 of MSMARCO with DeepImpact ranking. For k = 10, the MUF returned by $Q_k^4$ is 0.885, which is not shown in the figure for better readability.}
\label{fig:vary_prefix_size_deepimpact_bm25}
\vspace{-4mm}
\end{figure}

In Figure \ref{fig:vary_prefix_size_deepimpact_bm25} we see the results for DeepImpact. For simplicity, we only show two prefix configurations, huge and medium. Overall, we see somewhat similar behavior, with significant benefits over $Q^4_k$ for $k=10$ and $k=100$, little benefit for $k=1000$ without sampling, and good benefits once sampling is added again. We also observed similar results for QLD and DocT5Query; details are omitted due to space constraints.

{\bf Running Times:} 
Finally, we look at the computational cost of our threshold estimation techniques, and the reductions in computational costs obtained by using our threshold estimates in the MaxScore algorithm, one of the most widely used safe algorithms for top-$k$ disjunctive query processing.

        

\begin{table}[htbp]
    \vspace{-5mm}
    \caption{Running time of our estimation algorithm in nanosecond, on Clueweb09B with BM25 ranking.}
    \centering
    \vspace{-2mm}
    \scriptsize
    \begin{tabular}{|c|c|c|c|c|}
    \hline
             & \textit{ab} = 200 & \textit{ab} = 500 & \textit{ab} = 1k & \textit{ab} = 2k\\ 
            \hline
            Total (ns) & 52931 & 71228 & 101599 & 121436\\
            \hline
             Lookup Time & 10737 & 17962 & 25461 & 32760\\
            \# Lookup Made & 240 & 541 & 908 & 1361 \\
            Time / Lookup (ns) & 45 & 33 & 28 & 24\\
             \hline
        
    \hline
    \end{tabular}
    \label{tab:speed}
    \vspace{-3mm}
\end{table}

We start by looking at the cost of threshold estimation, which depends on the access budget $ab$ as well as the lookup budget $lb$. We observed costs of accessing and processing a posting from a prefix of about $25$ ns (nanoseconds), and typical costs of about $20$ to $40$ ns per lookup into an inverted list, though the precise amount depends on parameters such as the length of the list and the number of lookups into each list. There are also some non-trivial costs for finding all existing prefixes for a query, initializing various data structures, sorting accumulators by docID for sorted index lookups, and selecting the final threshold estimate.

\begin{table}[htbp]
    \vspace{-5mm}
    \scriptsize
    \caption{Running time of our estimation algorithm in nanoseconds, on MSMARCO with DeepImpact ranking.}
        \vspace{-2mm}
    \centering
    \begin{tabular}{|c|c|c|c|c|}
    \hline
             & \textit{ab} = 500 & \textit{ab} = 1k & \textit{ab} = 2k & \textit{ab} = 3k\\ 
            \hline
            Total (ns) & 111440 & 170049 & 218884 & 251110\\
            \hline
            Lookup Time & 7928 & 11870 & 13620 & 15201\\
            \# Lookup Made & 316 & 453 & 642 & 784 \\
            Time / Lookup (ns) & 25 & 26 & 21 & 19\\
             \hline
        
    \hline
    \end{tabular}
    \label{tab:speed2}
    \vspace{-2mm}
\end{table}

Table~\ref{tab:speed} shows running times for different access budgets in ns, on a medium size prefix for ClueWeb09B under BM25 ranking. Table~\ref{tab:speed2} contains corresponding results for MSMARCO under DeepImpact ranking. Overall, we get running times of about 100 to 250 us per estimate, which is affordable for many applications. We also see that the cost is not actually dominated by random lookups, which is maybe somewhat surprising given that random index lookups are commonly considered expensive in the IR community. Lookups on ClueWeb09B are somewhat more expensive than for MSMARCO given that the much larger ClueWeb09B collection has longer inverted lists.

\begin{table}
    \vspace{-8mm}
    \scriptsize
    \caption{Average query processing time (QT) of the MaxScore algorithm using different MUFs on ClueWeb09B using BM25, and on MSMARCO with DeepImpact ranking.}
  \centering
  \vspace{-2mm}
  \begin{tabular}{cc|cc|cc}
  & & \multicolumn{2}{c|}{\textbf{BM25 on ClueWeb09B}} & 
  \multicolumn{2}{c}{\textbf{DeepImpact}} \\
     & & $Q_k$ & Our Method & $Q_k$ & Our Method\\
    \hline
    \multirow{2}{*}{\textbf{\textit{k} = 10}} & MUF & 0.917 & 0.986 & 0.885 & 0.952\\
    & QT ($\mu$s) & 7626 & 6833 & 3668 & 3415\\
    \hline
    \multirow{2}{*}{\textbf{\textit{k} = 100}} & MUF & 0.936 & 0.973 & 0.928 & 0.957\\
     & QT ($\mu$s) & 10024 & 9155 & 5490 & 5183\\
    \hline
  \end{tabular}
  \label{tab:speedup_table}
  \vspace{-6mm}
\end{table}

Next, we look at resulting speedups when running the MaxScore algorithms with our threshold estimates, as compared to the case of $Q^4_k$ estimates. In Table \ref{tab:speedup_table}, we see results for MaxScore for two different threshold estimates: An estimate provided by $Q^4_k$, and an estimate from our method, on ClueWeb09B with BM25 ranking, and MSMARCO with DeepImpact ranking. For ClueWeb09B, we use \textit{ab} = 1k and \textit{lb} = 100 for \textit{k} = 10, and \textit{ab} = 5k and \textit{lb} = 500 for \textit{k} = 100. For Deepimpact, we use \textit{ab} = 500 and \textit{lb} = 500 for \textit{k} = 10, and \textit{ab} = 1k and \textit{lb} = 1k for \textit{k} = 100.

For the case of ClueWeb09B with BM25, we see significant decreases in running time that more than makeup for the increased running times of our threshold estimation techniques, with reductions by more than 800 us versus costs around 200us for threshold estimation. For MSMARCO with DeepImpact, things are a little closer, with reductions by about 250 to 300 us versus threshold estimation costs of about 110 us, but there is still some benefit. Overall, we see that our methods have the potential to decrease the running time of the MaxScore algorithms by providing more accurate threshold estimates.
\section{Conclusions and Future Work}

In this paper, we proposed top-\textit{k} threshold estimation methods that outperform the state-of-the-art quantile methods. Our results significantly narrow the gap between the state of the art and the ideal MUF of 1.0, at a moderate increase in computing and space overhead. Our methods work particularly well for small values of $k$, and for longer queries where quantile methods tend to not do well. We also evaluate our methods on four different ranking functions, including functions arising in recently proposed learned sparse indexing approaches.

\bibliographystyle{IEEEtran}
\bibliography{threshold}

\end{document}